\documentclass[9pt]{article}
\usepackage{textcomp}
\usepackage[utf8]{inputenc} 
\usepackage[T1]{fontenc}    
\usepackage{hyperref}       
\usepackage{url}            
\usepackage{booktabs}       
\usepackage{amsfonts}       
\usepackage{nicefrac}       
\usepackage{microtype}      
\usepackage{lipsum}		
\usepackage{graphicx}
\usepackage{doi}
\usepackage{biblatex}
\usepackage[utf8]{inputenc}
\usepackage{amsmath}

\usepackage{caption}
\title{Simple approach for extending the ambiguity-free-range of dual-comb ranging}

\begin{document}
\newcommand{\RomanNumeralCaps}[1]
    {\MakeUppercase{\romannumeral #1}}
\maketitle   
\noindent\textbf{Jakob Fellinger$^{1,*}$, Georg Winkler$^1$, P. E. Collin Aldia$^1$, Aline S. Mayer$^1$, Valentina Shumakova$^1$, Lukas W. Perner$^1$, Vito F. Pecile$^1$, Tadeusz Martynkien$^2$, Pawel Mergo$^3$, Grzegorz Soboń$^4$ and Oliver H. Heckl$^1$
}\vspace{5 mm}

\small{
\textit{
\centering{
$^1$University of Vienna, Faculty of Physics, Faculty Center for Nano Structure Research, Christian Doppler Laboratory for Mid-IR Spectroscopy and Semiconductor Optics, Boltzmanngasse 5, 1090 Vienna, Austria\\
$^2$Faculty of Fundamental Problems of Technology, Wrocław University of Science and Technology, Wybrzeze Wyspianskiego 27, 50-370 Wroclaw, Poland\\
$^3$Laboratory of Optical Fiber Technology, Maria Curie-Skłodowska University, pl. M. Curie-Sklodowskiej 3 , 20-031 Lublin, Poland\\
$^4$Laser \& Fiber Electronics Group, Wroclaw University of Technology, 27 Wybrzeże Wyspiańskiego Str., 50-370 Wroclaw, Poland\\
$^*$Corresponding author: jakob.fellinger@univie.ac.at\\
}
}
}
\vspace{10mm}

\begin{abstract}
\noindent Dual-comb (DC) ranging is an established method for high-precision and high-accuracy distance measurements. It is, however, restricted by an inherent length ambiguity and the requirement for complex control loops for comb stabilization. Here, we present a simple approach for expanding the ambiguity-free measurement length of dual-comb ranging by exploiting the intrinsic intensity modulation of a single-cavity dual-color DC for simultaneous time-of-flight and DC distance measurements. This measurement approach enables the measurement of distances up to several hundred km with the precision and accuracy of a dual-comb interferometric setup while providing a high data acquisition rate ($\approx$ 2 kHz) and requiring only the repetition rate of one of the combs to be stabilized. 
\end{abstract}
\vspace{10mm}

\noindent Recent progress in automation and metrology in various fields of application such as industrial robotics or autonomous transportation has led to an increasing demand for precise, fast, robust, and inexpensive ranging technologies which are applicable outside the laboratory frame.

Interferometric distance-measurement techniques based on homodyne or heterodyne detection offer high precision~\cite{Bobroff_1993}. In their most basic form, however, they require continuous tracking and unidirectional motion of the target to clearly interpret relative phase changes during the measurement. Otherwise, absolute measurements can only be obtained modulo the ambiguity-free range given by $\lambda/2$, where $\lambda$ is the optical wavelength. Alternatively, these ambiguity limitations can be overcome by using modulated or pulsed light. In these so-called time-of-flight (TOF) measurements the distance is measured by analyzing the temporal delay between the pulse coming back from the target and the pulse that has travelled along a known reference path~\cite{HICKMAN196947,Fortier2019}. However, such systems typically lack the high resolution of an interferometric system. 

Since their invention in the late 1990s, optical frequency combs (OFCs) have become an indispensable tool in optical metrology~\cite{Newbury2011,Fortier2019}. Initially used for counting the cycles of optical clocks, they quickly revolutionized many other fields, including applications in spectroscopy and ranging~\cite{Jang2018}. Various ranging methods based on OFCs have been reported so far, including optical cross-correlation~\cite{Lee2010}, synthetic-wavelength interferometry~\cite{Minoshima:00}, and multiple wavelengths referenced to a stabilized comb~\cite{Wang:15}, among others. 

Dual-comb ranging systems~\cite{Coddington2009} enable an ambiguity-free measurement range of $c/f_\mathrm{rep}$ (typically a few meters) in a single measurement via analysis of the beating signal between two optical frequency combs with slightly different repetition rates, i.e. different frequency spacings of the comb lines. A first comb is sent along a known reference path and along a signal path. At the end of both paths, the light is reflected back. The signal and reference reflections are sampled with a second comb. By evaluating the temporal shift of the respective interferograms the precision of the established TOF technique is improved by a factor of $\frac{f_\mathrm{rep,2}}{\Delta f_\mathrm{rep}}$~\cite{Coddington2009}. Here, $f_\mathrm{rep,2}$ is the repetition rate of the comb used for the remote sensing and $\Delta f_\mathrm{rep}$ is the difference to the repetition rate $f_\mathrm{rep,1}$ of the local reference comb. As these systems depend on a complex light source (two phase-stable frequency combs), different methods have been presented to simplify them while further improving their accuracy and precision ~\cite{Zhou:19,Zhu:19,8103879}. Via additional evaluation of the interferogram carrier phases, for example, it is possible to achieve wavelength resolution. A detailed review on the DC ranging technique is provided by Zhu et al. ~\cite{Zhu2018}.  

Different approaches have been reported to overcome the remaining ambiguity limitation, e.g. by using the Vernier-effect~\cite{Coddington2009}. The Vernier-effect method can be implemented by switching the role of the signal and reference path~\cite{Coddington2009}, by adjusting the repetition rate of the signal~\cite{Zhang:14} or by adding an extra measurement path~\cite{Liua:20}. However, all these approaches either introduce moving parts~\cite{Coddington2009}, or increase the measurement time and experimental complexity. Besides the Vernier-effect, one can exploit nonlinear optics to enable absolute distance ranging~\cite{7101801}, however, this method demands high peak powers.

An alternative high-precision ranging technique relies on using a single frequency comb and an electro-optical modulator (EOM)~\cite{Li:20}. This elegant method offers high-precision and long-distance measurement, but depends on the detection of weak side-bands, requiring high reflectivity of the target. 

Liu et al. used a single-cavity dual-comb~\cite{Liua:20} to reduce the complexity of the light source. However, they could only overcome the ambiguity limitation by exploiting the Vernier-effect.
 
This letter presents a new approach to extend the ambiguity-free range of DC ranging based on a single-cavity dual-color DC~\cite{Fellinger2019b} without the need to use the Vernier-effect or nonlinear optics. The ambiguity-free range is increased to $c/\Delta f_\mathrm{rep}$ (150~km for the laser parameters used in the presented experiment) within a single measurement, while allowing for data acquisition in the millisecond range.

Pulse trains emitted by single-cavity dual-color lasers are known to exhibit intensity modulations with a periodicity of $\Delta f_\mathrm{rep}$ caused by intracavity pulse collisions~\cite{Wei2018,Liu2020}. We exploit this periodic modulation imprinted on the individual pulse trains as a marker to perform a secondary long-range TOF measurement, which allows us to extend the non-ambiguity range. Additionally, we use the shift of the dual-comb interferograms to measure the distance with high precision.  

\begin{figure}[htbp]
\centering
\includegraphics[width=0.8\linewidth]{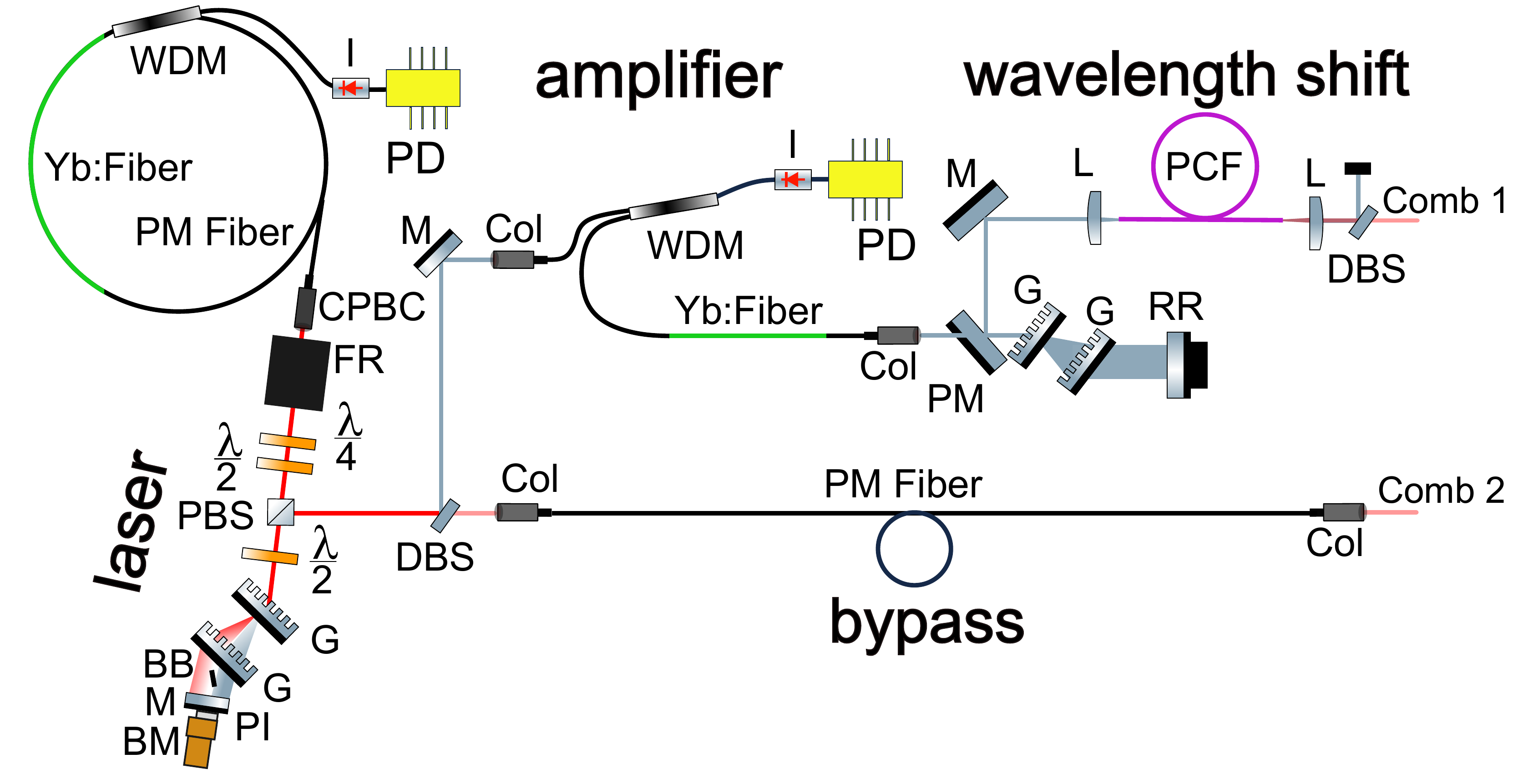}
\caption{92-MHz figure-9 dual-color dual-comb. WDM: wavelength division multiplexer, I: Isolator, CPBS: collimating polarisation beam combiner, FR: Faraday rotator, PBS: polarizing beam splitter, G: grating, M: mirror, $\frac{\lambda}{2}$: half wave palte, $\frac{\lambda}{4}$: quarter wave plate, PI: Piezos, BM: bullet mount, PD: pump diode, Col: collimator, L: Lens, PCF: photonic crystal fiber, RR: retroreflector, DBS: dichroic beam splitter, PM: D-shaped mirror.}
\label{fig:dc-laser}
\rule{\linewidth}{0.5pt}
\end{figure}
The system presented in this letter is based on an all-polarization-maintaining fiber laser using Ytterbium as a gain material and a nonlinear amplifying loop mirror in combination with a non-reciprocal phase bias for  mode-locking~\cite{Kuse2016,Hansel2017}. For a detailed analysis of how to operate and mode-lock these lasers, see~\cite{Mayer2020}. A sketch of the dual-color dual-comb can be seen in Fig.~\ref{fig:dc-laser}. 

\begin{figure}[htbp]
\centering
\includegraphics[width=0.8\linewidth]{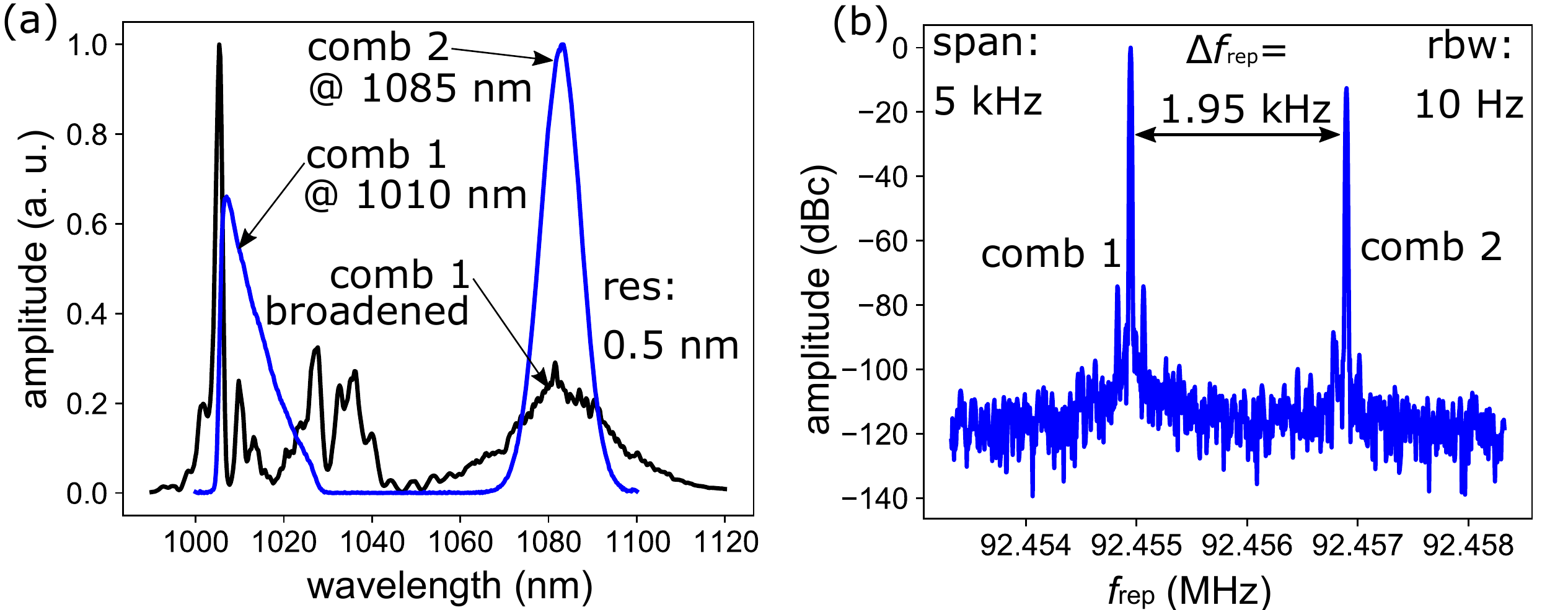}
\caption{92-MHz figure-9 dual-color laser state with one pulse operating in the anomalous dispersion regime (centred around 1085~nm) and one pulse in the normal dispersion regime centred around 1010~nm. (a) Optical output spectrum of the laser (blue) and spectrum of comb 1 after amplification and nonlinear broadening (black). (b) Zoom into the repetition rates of the two pulse trains.}
\label{fig:laser-state}
\rule{\linewidth}{0.5pt}
\end{figure}

The laser source used for this experiment is similar to the system presented in~\cite{Fellinger2019b}, however, the repetition rate of the laser is increased to around 92~MHz. A tuneable mechanical spectral filter~\cite{Fellinger2019a} allows the all-PM laser to emit two independent pulse trains with different center wavelengths and slightly different repetition rates. As shown in ~\cite{Fellinger2019b}, the two pulse trains may run in a different net intracavity dispersion regime each. The output spectrum of the laser is measured using an optical spectrum analyzer (Ando AQ6315A), see Fig.~\ref{fig:laser-state}(a). One of the two pulse trains has a center wavelength of 1010~nm (running in the normal dispersion regime), an output power of 1.5~mW and a repetition rate of 92.455~MHz, while the other one has a center wavelength of 1085~nm (running in the anomalous dispersion regime), an output power of 2~mW and a repetition rate of 92.457~MHz. Additionally, we analyzed the laser output using a fast photodiode (Thorlabs DET08CL/M) connected to a microwave spectrum analyzer (Keysight PXA N9030B). Figure~\ref{fig:laser-state}(b) shows a zoom into the two repetition rates of the two pulse trains operating inside the single laser cavity with $\Delta f_\mathrm{rep}\approx~$1.95~kHz. The large spectral separation of $\approx 75$~nm enables a clean spectral separation of two combs at the output of laser cavity by using a standard dichroic mirror with a cut-on wavelength of 1050~nm (Edmund Optics \#87-041). 

 We generate a spectral overlap between the two combs and thus enable a dual-comb beating signal via amplification and nonlinear broadening: first, we amplify comb~1 inside a PM-Yb single-mode fiber-amplifier to approximately 30~mW. The output of the amplifier is temporally compressed using a grating compressor consisting of two gratings ($\RomanNumeralCaps{2}- \RomanNumeralCaps{6}$ Transmission Grating: T-1000-1040-31.8x12.3-94) used in double-pass configuration. The temporally compressed ($\approx 250$~fs) light is coupled into a highly nonlinear fiber, similar to the one used in \cite{Sobon:17}, with a coupling efficiency of $\approx $ 65\%. Inside the fiber, comb~1 is spectrally broadened to overlap with comb~2 (see Fig.~\ref{fig:laser-state}(c)) and subsequently sent through a long-pass filter with a cut-on wavelength of 1050~nm (Thorlabs FELH1050).

\begin{figure}[t]
\centering
\includegraphics[width=0.8\linewidth]{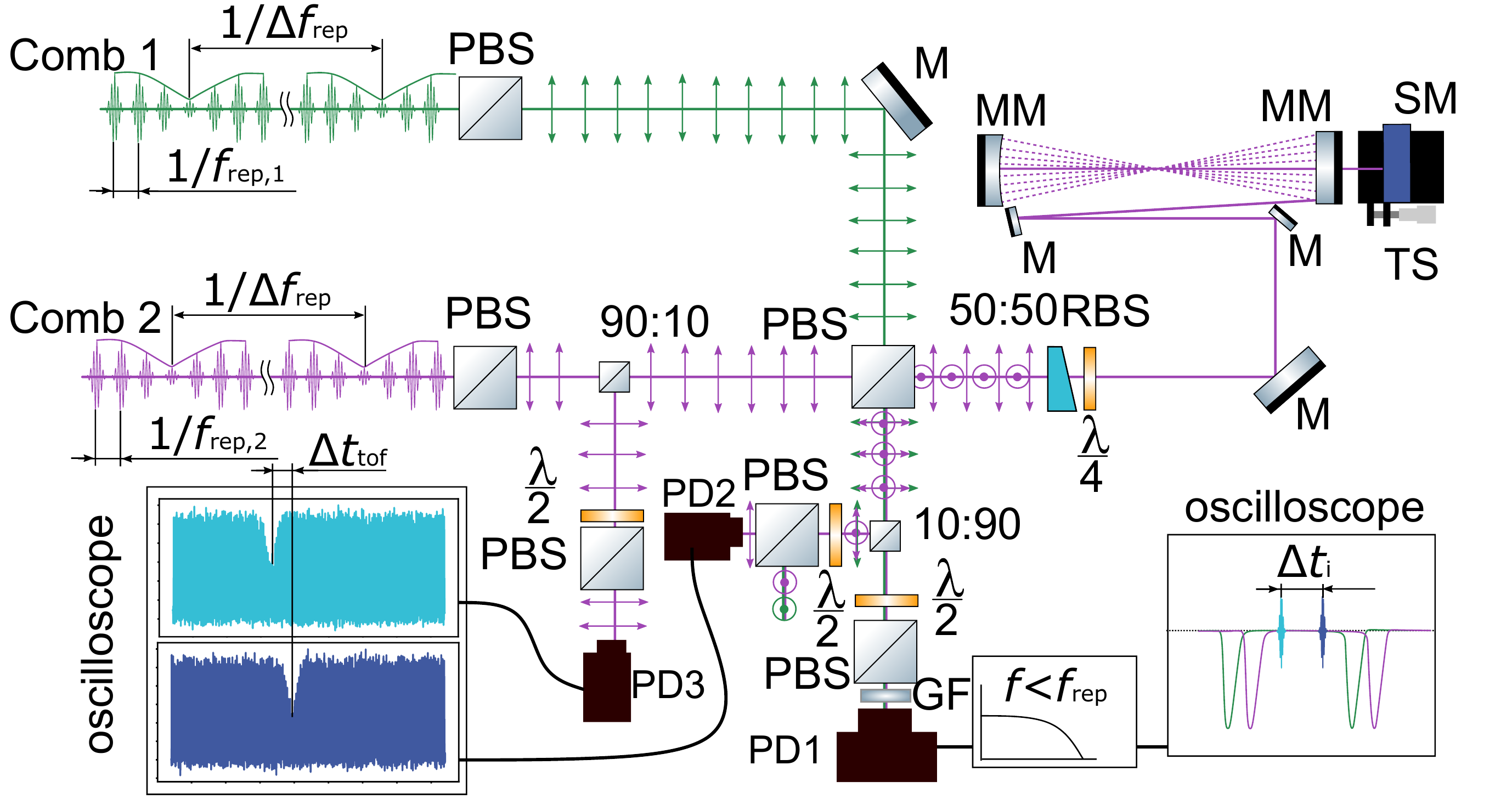}
\caption{Working principle of dual-comb ranging based on a single-cavity dual-comb. The single cavity is emitting two intensity-modulated frequency combs (comb~1 and comb~2) with slightly different repetition rates ($\Delta f_\mathrm{rep} \approx$ 2 kHz) around 92 MHz. Comb~2 is used as the remote sensing probe. Comb~1 is used to sample comb~2 for precise dual-comb ranging by evaluating the time-delay $\Delta t_\mathrm{i}$ between the reference and signal interferogram recorded on photodiode 1 (PD1). Photodiode 2 (PD2) and photodiode 3 (PD3) provide the signals to determine the time-of-flight delay $\Delta t_\mathrm{tof}$ used for coarse length measurement. PBS: polarizing beam splitter, M: mirror, $\lambda/2$: half wave plate, $\lambda/4$: quarter wave palate, 90:10: beam splitter, 50:50RBS: reference beam splitter, GF: optical band pass filter, TS: translation stage, MM: multi-pass cell mirror, SM: signal mirror.}
\label{fig:messprinz}
\rule{\linewidth}{0.5pt}
\end{figure}

\begin{figure}[htbp]
\centering
\includegraphics[width=0.8\linewidth]{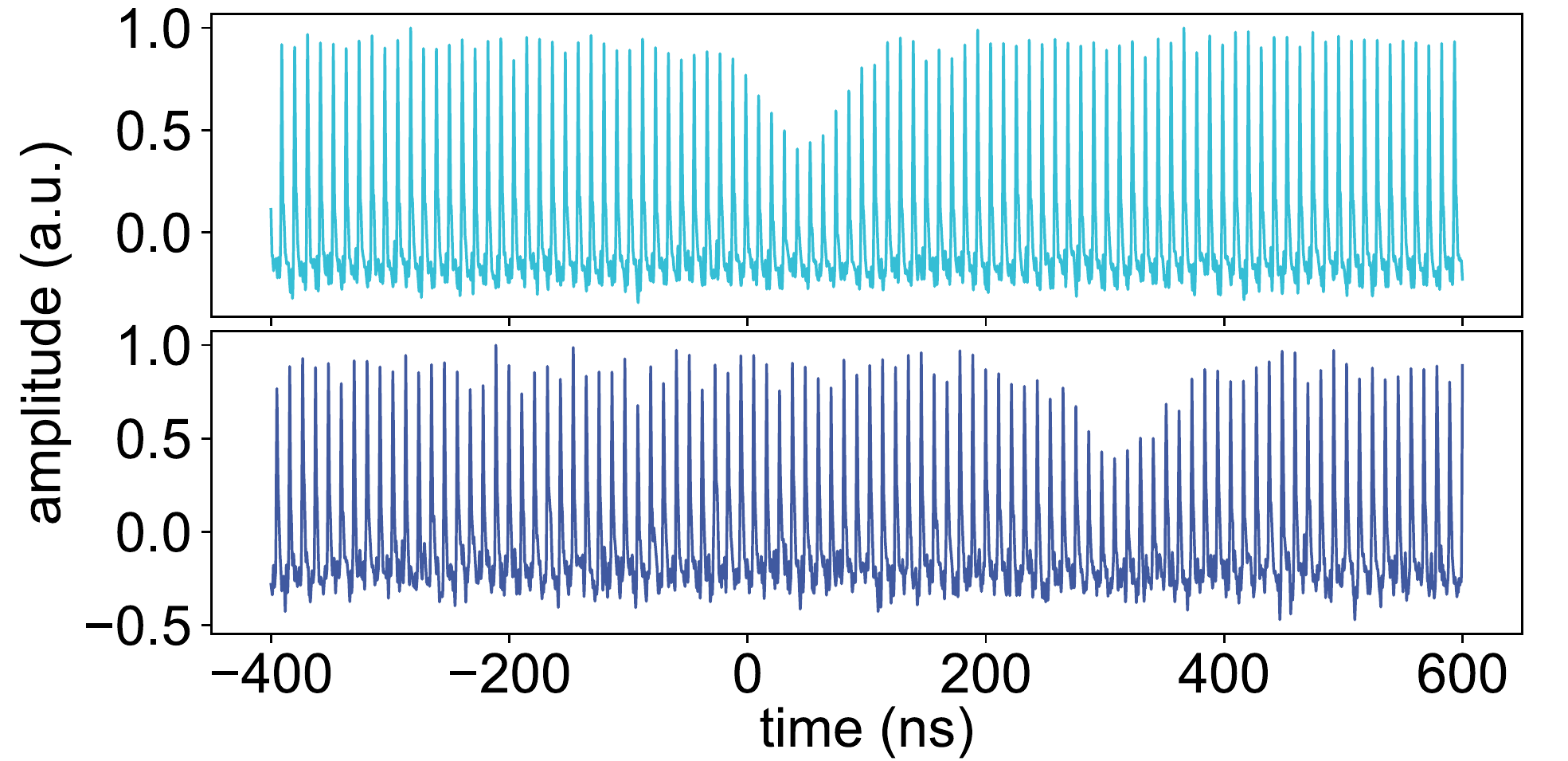}
\caption{Time-of-flight measurement using the intensity modulation on the pulse train as a marker. The turquoise trace (top) shows the reference measured at PD3 (Fig.~\ref{fig:messprinz}). The dark blue trace (bottom) shows the signal measured at PD2 (Fig.~\ref{fig:messprinz}). Due to the path length difference, the modulation is shifted. The time-delay between the modulation signals is used for coarse length calculation.}
\label{fig:TOF}
\rule{\linewidth}{0.5pt}
\end{figure}

A detailed sketch of the complete ranging experiment can be seen in Fig.~\ref{fig:messprinz}.
The cross-phase modulation signal is imprinted on both combs and is visible as an intensity modulation on both pulse trains, see Fig.~\ref{fig:TOF}. For the TOF measurement, only comb~2 is required. In our implementation, the distance $L$ to be measured is the path delimited by the reflective side of the partial reflector (50:50RBS, turquoise optic in Fig~\ref{fig:messprinz}) and the reflective side of the end-mirror (SM, blue optic in Fig~\ref{fig:messprinz}).

The light reflected by the 50:50RBS (starting point of $L$), is sent onto a fast photodiode (PD3) (Thorlabs DET08CL/M) and used as a reference. The light reflected back by the end-mirror is sent onto a second fast photodiode (PD2, Thorlabs DET08CL/M). The photodiodes are connected to an oscilloscope (Teledyne LeCroy WavePro 7Zi 6~GHz-40GS/s) to evaluate the time traces. The length is evaluated by measuring the temporal shift of the cross-phase modulation signal between the reference (PD3) and the signal (PD2). A typical measurement is shown in Fig.~\ref{fig:TOF}. Note that the detection of the two beams on different photodiodes is not strictly required, but simplifies the analysis of the two pulse trains by separating them directly. However, to compensate for the different distances of the two photodiodes from the reflector 50:50RBS, a seperate TOF calibration measurement needs to be performed. This calibration is easily implemented by sending the reference beam onto both photodiodes. The accuracy of the TOF measurement must only be on the order of a single cavity length ($\approx3.2$~m), which is easily achieved. 

Taking into account the calibration and the bandwidth of the two photodiodes (5~GHz), we achieve a precision of a few millimeters with the TOF measurement. The ambiguity-free length for this measurement is given by  $c/\Delta f_\mathrm{rep}$, which corresponds to 150~km for $\Delta f_\mathrm{rep}=2$~kHz. Hence, we use the TOF measurement to calculate the integer division factor $n=\mathrm{int}\left(\frac{L}{L_\mathrm{c}}\right)$ where  $L_\mathrm{c}=c/f_\mathrm{rep,2}$ is the optical cavity length.

To additionally measure the distance with high precision, we simultaneously implement the dual-comb ranging-principle~\cite{Coddington2009}. We use comb~1 to sample the signal beam (coming back from the end mirror, blue optic) as well as the reference beam (coming back from the partial reflector, turquoise optic) and measure the optical beating using photodiode PD1 (Thorlabs PDA05CF2). The fact that the reference beam is also able to reach PD1 despite being vertically polarized is due to the low extinction ratio of the PBS when used in reflection. To avoid optical aliasing, we filter the light in front of PD1 using a spectral filter (a grating and slit) and focus it on the photodiode.    

\begin{figure}[htbp]
\centering
\includegraphics[width=0.8\linewidth]{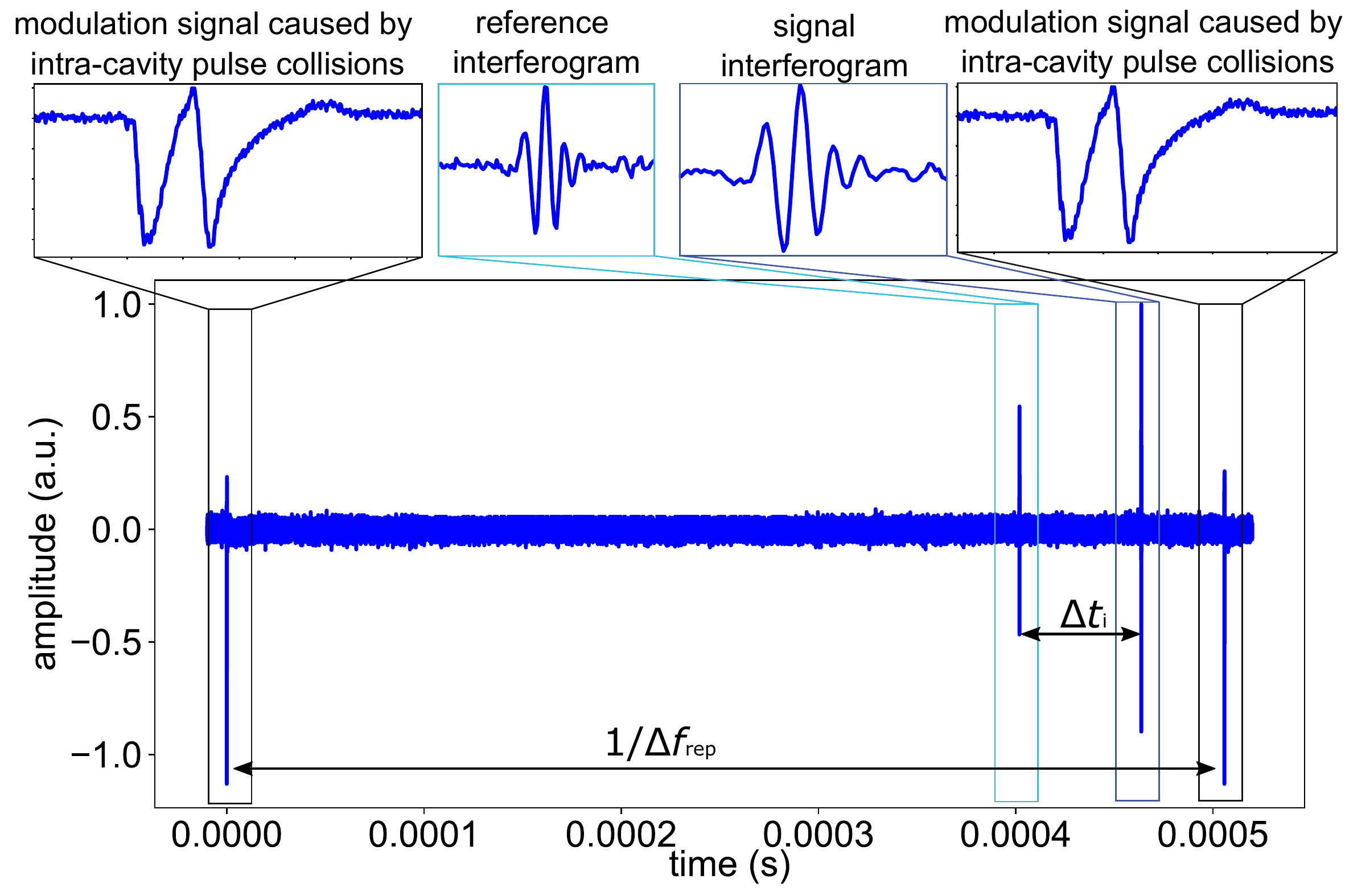}
\caption{Dual-comb ranging measurement based on a signal trace acquired via PD1. A complete scan of 1/$ \Delta f_\mathrm{rep}$ is shown, corresponding to the minimum measurement time.}
\label{fig:messtrace}
\rule{\linewidth}{0.5pt}
\end{figure}

The signal of PD1 is electronically filtered with a 70 MHz low pass filter (to filter away the repetition rate) and analyzed using an oscilloscope. Looking at the filtered time trace, we get four signals, each signal occurring periodically with a repetition frequency of $\Delta f_\mathrm{rep}$. The signals correspond to the modulation signals occurring on pulse train~1 and pulse train~2, the dual-comb interferogram corresponding to the reference beam, and the dual-comb interferogram corresponding to the signal beam. A typical measurement trace is shown in Fig.~\ref{fig:messtrace}. 

Since PD1 is an amplified detector, power levels of a few \textmu W for the reference and signal beam are sufficient to measure the interferograms. However, PD2 and PD3 are biased detectors (low cost and high bandwidth), requiring a power of around 50~\textmu W to achieve an adequate signal-to-noise ratio for evaluation of the time traces.  The signal-to-noise ratio could be improved by using amplified fast detectors. In this case, a few \textmu W signal would be sufficient to perform the ranging measurement, making it a promising candidate for applications outside the laboratory frame.    
The time-trace of PD1 is evaluated by measuring the time delay between the reference and signal interferogram. The time delay between the two interferograms is given by
\begin{equation}
\Delta t_\mathrm{i}= \frac{L_\mathrm{i}}{c}\frac{f_\mathrm{rep,2}}{\Delta f_\mathrm{rep}},
\label{eq:refname1}
\end{equation}
where $L_\mathrm{i}$ is the short-scale measurement length as defined below. $L_\mathrm{i}$ depends on the temporal position of the signal interferogram relative to the reference interferogram. To identify the reference and signal interferogram, we set different power levels, i.e. different amplitudes, of the interferogram. We must distinguish between three cases: 

\begin{itemize}
\item \textbf{Case 1}: The signal is delayed compared to the reference interferogram (this is the case for the time-trace plotted in Fig.~\ref{fig:messtrace}): 
$L_\mathrm{i}=(L\mod L_\mathrm{c})$.
\item \textbf{Case 2}: The reference is delayed compared to the signal interferogram:
$L_\mathrm{i}=L_\mathrm{c}-(L\mod L_\mathrm{c})$.
\item \textbf{Case 3}: The reference is exactly at the same position as the signal interferogram:
$L_\mathrm{i}=0$.
\end{itemize}

The total length of the distance measured is then given by:
\begin{equation}
L=n\cdot L_\mathrm{c}+L_\mathrm{i}.
\label{eq:refname2}
\end{equation}
To achieve sufficient accuracy, $f_\mathrm{rep,2}$ and $\Delta f_\mathrm{rep,2}$ can either be tracked or stabilized. Here, we stabilized $f_\mathrm{rep,2}$ (by stabilizing the 15th harmonic $f_\mathrm{rep,2}$ $\approx$ 1.386~GHz) to a frequency generator (Rohde\&Schwarz SMF 100A) using a piezo-controlled end mirror inside the laser resonator, and we tracked $\Delta f_\mathrm{rep}$ by measuring the periodicity of the cross-phase modulation signal in the time trace.

\begin{figure}[tb]
\centering
\includegraphics[width=0.8\linewidth]{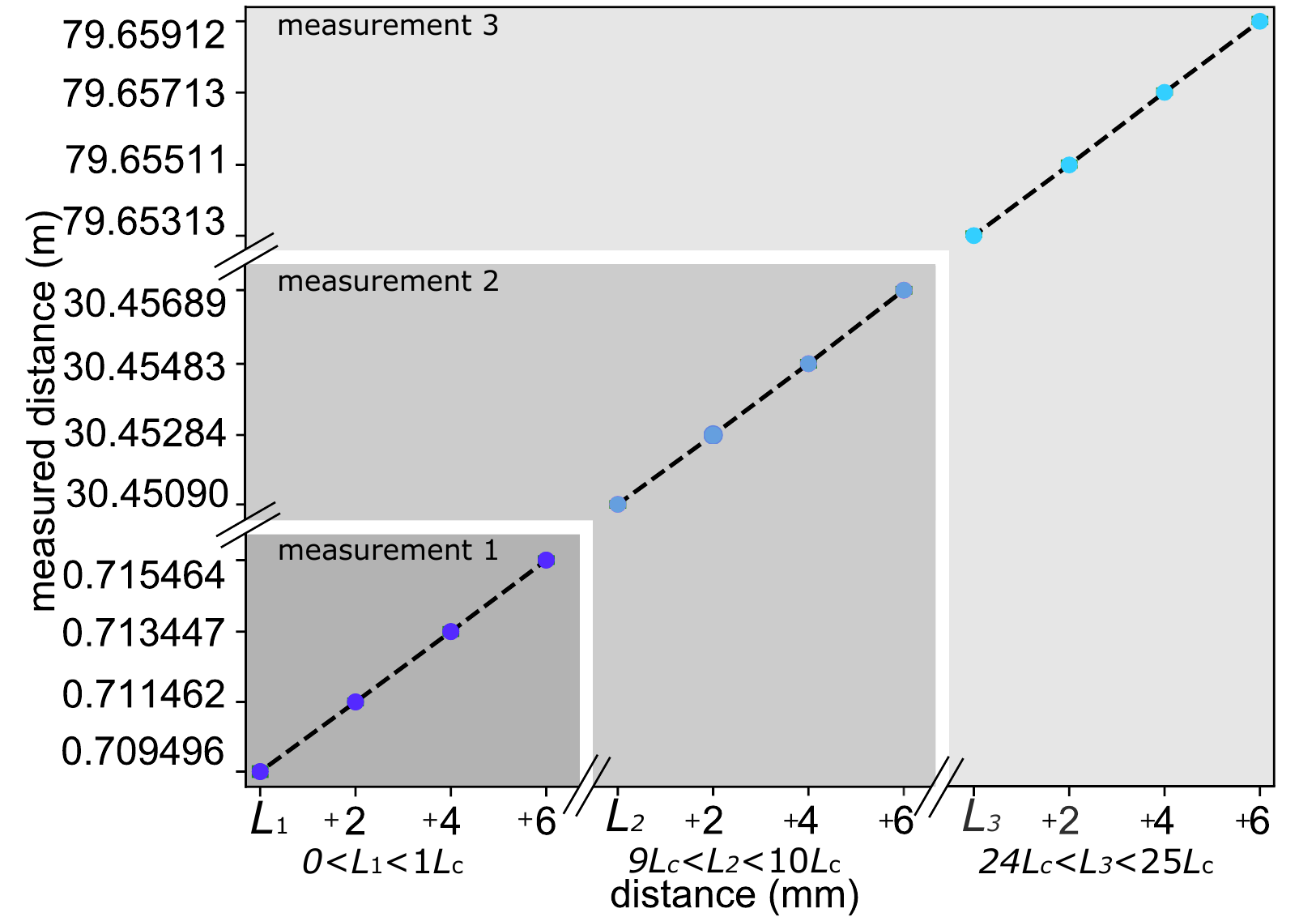}
\caption{Evaluation of the combined time-of-flight and dual-comb measurements. Using both signals, we are not limited by the ambiguity range of traditional dual-comb ranging. The dark blue section is measured at a starting distance $L_1$ within a single cavity length ($L_\mathrm{c}=c/f_\mathrm{rep}$), which is the maximum ambiguity range for conventional dual-comb-ranging. The light blue section is measured with a distance which is $L_2>9L_\mathrm{c}$ and the turquoise section is measured with a distance $L_3>24L_\mathrm{c}$. To verify the precision of the measurement, we changed the end-mirror position using a micrometer translation stage.}
\label{fig:finalmess}
\rule{\linewidth}{0.5pt}
\end{figure}

We verified the capability of our system to perform single-shot measurements over a large ambiguity-free range by measuring the path lengths of different multi-pass cell configurations. We varied the measurement distance $2L$ (to the end mirror and back) from less than a single cavity length (measurement 1 in Fig.~\ref{fig:finalmess}) to over more then 9 cavity lengths (measurement 2 in Fig.~\ref{fig:finalmess}) and up to more than 24 cavity lengths (measurement 3 in Fig.~\ref{fig:finalmess}). Using the presented approach, we were able to measure the different distances without any ambiguity in a single sweep (measurement time $=1/\Delta f_\mathrm{rep}= 500$~\textmu s). In each configuration we changed the path-length in increments of 2~mm (by scanning the end-mirror in 1-mm steps using a translation stage). For each measurement point, we took 15 measurements and got an error of $\sigma <100$~\textmu m. This precision is limited by the dual comb parameters. However, it can be significantly improved by increasing $f_\mathrm{rep}$ while decreasing $ \Delta f_\mathrm{rep}$ at the expense of an increase in measurement time.  

The presented measurement principle enables single-shot high-precision absolute-distance ranging over an ambiguity-free measurement range of around 150~km by combining the advantages of TOF measurements and dual-comb ranging. The system is based on a single light source with much less stringent requirements on stabilization (only $f_\mathrm{rep,2}$ needs to be either stabilized or continuously measured) compared to two mutually phase-stable (locked) frequency combs. In addition, the measurement does not rely on nonlinear processes. Hence, it only needs \textmu W-level reflected powers returning to the sensor. The high precision, large ambiguity-free measurement range and  fast data acquisition in combination with the high sensitivity, make the presented measurement approach an ideal system for field applications.
\vspace{5mm}

\textbf{Funding} Austrian Federal Ministry for Digital and Economic Affairs; National Foundation for Research, Technology, and Development; Christian Doppler Forschungsgesellschaft; Austrian Science Fund (FWF) (M2561-N36) and (P 33680 Einzelprojekte); Foundation For Polish Science - Fundacja na rzecz Nauki Polskiej, (FNP) (First TEAM/2017-4/39).

\vspace{5mm}
\textbf{Disclosures}
The authors declare no conflicts of interest.

\printbibliography
\end{document}